\documentclass[twocolumns]{aa}
\usepackage{graphicx}
\usepackage{epsfig}
\usepackage{textcomp}

\begin{document}

\title{A high-sensitivity 6.7 GHz methanol maser survey toward H$_{2}$O sources}
\author{Y. Xu\inst{1,2}, J. J. Li\inst{3}, K. Hachisuka\inst{3}, J. D.
Pandian\inst{1},  K. M. Menten\inst{1}, C. Henkel\inst{1}}

   \offprints{Y. Xu}
   \institute{
   Max-Planck-Institute f\"{u}r Radioastronomie,
Auf dem H\"{u}gel 69, 53121 Bonn, Germany\\
         \email{xuye@mpifr-bonn.mpg.de}
         \and
   Purple Mountain Observatory, Chinese Academy of Sciences, Nanjing 210008, China
         \and
   Shanghai Astronomical Observatory, Chinese Academy of Sciences, Shanghai 20030, China}

   \date{Received date; accepted date}

 \abstract{
 We present the results of a high sensitivity survey for 6.7 GHz methanol masers towards 22 GHz water maser using
 the 100 m Efflesberg telescope. A total of 89 sources were observed and 10 new methanol masers were detected.
 The new detections are relatively faint with peak flux densities between 0.5 and 4.0 Jy.
 A nil detection rate from low-mass star forming regions enhances the conclusion that the masers are
 only associated with massive star formation. Even the faintest methanol maser in our survey,
 with a luminosity of 1.1 $10^{-9} L_\odot$ is associated with massive stars as inferred from its infrared luminosity.


}
 \authorrunning{Y. Xu et al.}
 \titlerunning{Methanol maser survey}

\maketitle

\section{Introduction}
The $5_1-6_0$ A$^+$ transition of methanol at 6.7 GHz produces the
brightest known methanol masers. These masers are widespread in the
Galaxy and more than 550 sources have been detected to date,
including the compilations of Xu et al. (2003), Malyshev \& Sobolev
(2003) and Pestalozzi et al. (2005) and the new searches of Pandian
et al. (2007) and Ellingsen (2007). The masers are not only a
powerful tool of massive star-forming regions, but are also
potentially useful for measuring distances using VLBI techniques, as
has been demonstrated using their 12.2 GHz counterparts (Xu et al.
2006). Searches for 6.7 GHz methanol masers have been primarily
targeted toward IRAS sources, OH masers, and ultracompact HII
regions (e.g. Caswell et al. 1995; Walsh et al. 1997; Szymczak et
al. 2000). In addition there have been a few blind surveys (Caswell
et al. 1995; Ellingsen et al. 1996, Szymczak et al. 2002, Pandian et
al. 2007). Unlike surveys in Southern sky, most surveys in Northern
sky have a sensitivity limit of more than 1 Jy. The recent blind
survey using the 305 m Arecibo radio telescope resulted in the
discovery of numerous faint ($<$ 1 Jy) methanol masers (Pandian et
al. 2007).

Here, we report on the results of a sensitive survey for CH$_{3}$OH
masers targeted towards H$_{2}$O maser sources, primarily in the
Northern sky. Although there is no spatial correlation between 6.7
GHz CH$_{3}$OH and 22 GHz H$_{2}$O masers (Beuther et al. 2002;
Breen et al. 2007), a high detection rate of both maser types in the
same star forming regions indicates that the maser bearing phases
for these two species overlap (Codella \& Moscadelli 2000; Szymczak
et al. 2005). The goals of our survey are two-fold: to find more
CH$_{3}$OH masers as astrometric targets for future VLBI
observations to understand the spiral structure of the Galaxy.
Further, since H$_2$O masers are found towards both low-mass and
high-mass star forming regions, we hope to verify the exclusive
association of 6.7 GHz methanol masers with massive star formation.

\section{Observations}
The observations were made using the Effelsberg 100 m
telescope\footnote{Based on observations with the 100-m telescope of
the MPIfR (Max-Planck-Institut f\"{u}r Radioastronomie) at
Effelsberg.} in February and May 2006. The rest frequency adopted
for the $5_1-6_0$~A$^+$ transition was 6668.519 MHz (Breckenridge \&
Kukolich 1995). For the February observations, the spectrometer was
configured to have a 10 MHz bandwidth with 4096 spectral channels
yielding a spectral resolution of 0.11 km s$^{-1}$ and a velocity
coverage of 450 km s$^{-1}$. In May, a bandwidth of 20 MHz was used
giving a spectral resolution of 0.22 km s$^{-1}$ and a velocity
coverage of 900 km s$^{-1}$. The half-power beam width was $\sim$
2$'$ and the telescope has an rms pointing error of 10$''$. The
observations were made in position switched mode. The system
temperature was typically around 35 K during our observations. The
flux density scale was determined by observations of NGC7027 (Ott et
al. 1994). The absolute calibration for flux density is estimated to
be accurate to $\sim$ 10\%. The integration time on source was
typically three minutes, which resulted in a mean rms noise level of
$\sim$ 0.1 - 0.2 Jy in the spectra. When a source was detected, the
integration time was increased to around 8 minutes (at the same
position) with a velocity resolution of 0.11 km s$^{-1}$ to obtain
spectra with high signal to noise ratio.

The target sources were H$_2$O masers selected from the Arcetri
catalog (Comoretto et al. 1990 and Brand et al. 1994), and are shown
in Table 1. The sources were selected based on associations with
star forming regions or HII regions, with declinations $\delta \geq
-10^{\circ}$. This gave us a sample of 178 sources of which 17
sources may be associated with low-mass young stellar objects (YSOs)
with infrared luminosities less than $10^{3} L_{\odot}$. 154 out of
178 sources are associated with IRAS point sources. We then excluded
sources that had previous detections of methanol masers, which
reduced our sample size to 131 sources. Of this, we observed a total
of 89 sources within our observing time constraints. For all
sources, the spectrometer was centered on the velocity of peak
emission of the water maser. However, the wide velocity coverage
precludes the possibility of methanol masers being missed due to
their velocities being significantly offset from that of the water
masers.

\section{Results}
Our observations resulted in the discovery of 10 new methanol
masers, the properties of which are listed in Table 2. Since we did
not attempt to refine the position of the methanol masers using a
grid of observations, the positions quoted in Table 2 could have an
error as high as $\sim 1'$. The maser luminosities quoted in Table 2
are calculated from the integrated flux density assuming isotropic
emission. Details of water maser sources that had non-detections of
methanol masers are indicated in Table 3, which is available
on-line. The peak flux densities of the methanol masers detected in
our survey range from 0.5 to 4.0 Jy. Five sources are located beyond
the solar circle, which is a significant addition to the number of
such sources in the northern sky. It is interesting that the
kinematic distance to the source 05137+3919 puts it at a distance of
$\sim$ 14 kpc from the Sun and 20 kpc from the Galactic center. This
is one of the farthest methanol masers (in terms of distance from
the Galactic center) in the outer Galaxy, although the uncertainties
in the rotation curve at these galactrocentric radii, and peculiar
motions such as that observed in W3OH (Xu et al. 2006) translate to
significant uncertainties in the kinematic distance. Measuring
parallax distances to sources like 05137+3919 will be useful for
measuring the rotation speed of the Galaxy at large galactocentric
radii.

\subsection{Individual sources}
The spectra of the CH$_{3}$OH masers detected in our survey are
shown in Figure 1. The spectra have a velocity resolution of 0.11 km
s$^{-1}$. Here we present notes on individual sources.

\noindent \begin{bf}{05137+3919}\end{bf}. \ There are two features
that are separated by about 4.5 km s$^{-1}$. The stronger feature is
at an LSR velocity of --3.9 km s$^{-1}$, while the weaker feature
has a flux density of only 0.35 Jy. This region is associated with a
3.6 cm continuum source (Molinari et al. 2002).

\noindent \begin{bf}{06446+0029}\end{bf}. \ There are at least five
features over a velocity range of over 7 km s$^{-1}$. The feature at
48.6 km s$^{-1}$ is the strongest one. There is near infrared
emission in this region as seen in 2MASS, and a non-detection of SiO
masers (Harju et al. 1998).

\noindent \begin{bf}{18319-0802}\end{bf}. \ There are several
features spanning a velocity range over 15 km s$^{-1}$. The weakest
feature is only about 0.2 Jy. An ultracompact (UC) HII region,
separated by about 40$''$, could be associated with this region
(Becker et al. 1994).

\noindent \begin{bf}{18355-0650}\end{bf}. \ There are at least five
features spanning a velocity range of about 6 km s$^{-1}$. An UC HII
region, separated by 17$''$, is associated with this region (Becker
et al. 1994).

\noindent \begin{bf}{G29.91-0.05}\end{bf} Features span from 93.4 to
105.3 km s$^{-1}$ with multiple features being blended together.
There are at least two compact HII regions associated with this
region within 2$'$ (Wood \& Churchwell 1989a; Becker et al. 1994).
One of them, separated by about 80$''$, could be associated with
this region.

\noindent \begin{bf}{18403-0440}\end{bf}. \ This source primarily
shows a single feature at +20.1 km s$^{-1}$. No observations have
been reported on this region, except near infrared emission from the
2MASS.

\noindent \begin{bf}{18479-0005}\end{bf}. Walsh et al. (1997) did
not detect maser emission within their 1 Jy limit (3$\sigma$). There
are several features crowded within a velocity range of only 2 km
s$^{-1}$. A UC HII region is associated with this source (Kurtz et
al. 1994).

\noindent \begin{bf}{20275+4001}\end{bf}. \ This source displays a
single feature at -6.9 km s$^{-1}$, which matches the
H$^{13}$CO$^{+}$ (4-3) peak of Hasegawa \& Mitchell (1995). A
bipolar outflow and a continuum source were also detected in this
region (Hasegawa \& Mitchell 1995; Trinidad et al. 2003).
Mid-infrared images show that this source is surrounded by an
optically thick dusty envelope (Marengo et al. 2000).

\noindent \begin{bf}{21306+5540}\end{bf}. \ There are clearly five
features in this source. Except for near infrared emission from the
2MASS, no other observations have reported in this region.

\noindent \begin{bf}{22176+6303 (S140)}\end{bf}. \ This source shows
a double peaked structure and is the weakest source detected in this
survey. At a distance of 910 pc (Crampton \& Fisher 1974), the maser
luminosity is very low (1.1 10$^{-9} L_{\odot}$). However, its
infrared luminosity indicates that it is still associated with a
massive star forming region (Table 2). A faint continuum source
(Kurtz et al. 1994) and a CO outflow (Minchin et al. 1993) are
associated with the region.

\section{Discussion}

We detected 10 methanol masers from targeting 89 sources which
results in a detection rate of $\sim$ 11\%. However, to compare the
statistics of H$_2$O and methanol masers, we have to consider the
entire original sample of 178 sources that satisfied our selection
criteria ($\delta \geq -10^{\circ}$). 47 sources have previous
detections of methanol masers, while 42 sources were not observed
due to constraints of observing time. Hence, the overall detection
rate of methanol masers in a water maser sample is at least $\sim$
32\%. Our sample includes 17 sources that are associated with
low-mass YSOs and 10 of them were observed in this survey. A nil
detection rate from these sources adds to the results of Minier et
al. (2003), suggesting that 6.7 GHz methanol masers are only
associated with massive star forming regions. Therefore, excluding
the low-mass YSOs, the detection rate is at least $\sim$ 35\%.

Fig. 2 shows the color-color diagram of the IRAS sources associated
with the sample of water masers. The box on the upper right corner
shows the Wood \& Churchwell (1989b) (hereafter WC) criteria used to
identify embedded massive stars and ultracompact HII regions. The
stars and circles show the old and new methanol maser detections
respectively. 113 out of 154 sources satisfy the WC criteria, of
which 40 sources have methanol maser emission. Since 21 sources
satisfying the WC criteria were not observed, the detection rate
among IRAS sources satisfying WC criteria and hosting H$_2$O masers
is at least $\sim$ 35\%. This detection rate is better than that of
a survey based purely on IRAS sources satisfying WC criteria (e.g.
see discussion of Ellingsen et al. 1996). It is to be cautioned that
not all of these cases are true associations, as the IRAS source may
point to the brightest far infrared source in the star forming
region, and better positions for methanol masers may preclude some
of the current associations (Ellingsen 2006; Pandian \& Goldsmith
2007). There are also 12 methanol masers whose IRAS colors do not
satisfy WC criteria. It can also be seen from Fig. 2 that there is
no distinction between the IRAS sources associated with the new
detections, and those associated with the previous detections (which
are brighter), nor is there any distinction between the colors of
sources with and without methanol masers. Thus, the low detection
rate in our survey (11 \%) is not due to any systematic differences
between in the infrared properties of the sources in our survey
(versus the sources associated with previous detections of methanol
masers). The lack of distinction of the IRAS source properties of
bright versus faint methanol masers is also consistent with the
observation of Pandian \& Goldsmith (2007).

There is only one detection with a peak flux density less than 1 Jy,
(this is also the only such source in the entire sample including
previous detections). It should be noted that our sample is not from
an unbiased survey, and hence it is not possible to discuss the
implications of this in the context of the methanol maser population
in the Galaxy. However, we note that this is consistent with the
results of the simulation of van der Walt (2005) and the results of
the Pandian et al. (2007). van der Walt (2005) using Monte Carlo
simulations determined the completeness of surveys as a function of
their flux densities, while the observational results of Pandian et
al. (2007) corroborate the theoretical analysis with regard to the
total number of methanol masers in the Galaxy. Moreover, Pandian et
al. (2007) found that the distribution of peak flux densities drops
at flux densities below $\sim$ 1 Jy. This could be one of the
reasons why we didn't detect many sources below 1 Jy. Fig. 3a (left
panel) shows the luminosity distributions for both maser types as a
function of the infrared luminosity of the host IRAS sources and
Fig. 3b (right panel) shows the the maser luminosities of the two
species plotted against each other. Only sources that show emission
in both species are shown in Fig. 3. The infrared flux $F_{IR}$
(used to determine the infrared luminosity $L_{IR}$) is calculated
using the formula below (Casoli et al. 1986).
\begin{equation}
F_{IR} (10^{-13} \rm{W m}^{-2}) = 1.75(\frac{F_{12}}{0.79} + \frac{F_{25}}{2} + \frac{F_{60}}{3.9} + \frac{F_{100}}{9.9})
\end{equation}
where $F_{12}$, $F_{25}$, $F_{60}$ and $F_{100}$ refer to the IRAS
fluxes in 12, 25, 60 and 100 $\mu$m respectively. The distances are
taken from the literature. For sources with no published distance,
the near kinematic distance, computed from the peak velocity of 6.7
GHz emission using the galactic rotation curve of Wouterloot \&
Brand (1989), assuming $R_{0}$ = 8.5 kpc and $\Theta_{0}$ = 220 km
s$^{-1}$, is adopted. The near kinematic distance is used because it
seems realistic (Sobolev et al. 2005).

Fig. 3a shows that there is reasonably good correlation seen between
the infrared luminosity and that of both H$_{2}$O (Correlation
coefficient R = 0.64, Probability $p <$ 0.0001) and CH$_{3}$OH
masers (R = 0.56, $p <$ 0.0001). This correlation has been found in
the past by a number of groups (e.g. Wouterloot \& Walmsley 1986;
Szymczak et al. 2005), and the vertical scatter is normally
attributed to the variability of the masers. However, it is not
clear as to whether this correlation is physically meaningful. On
the one hand, both maser types are spatially separated and have very
different excitation requirements. H$_{2}$O masers are collisionally
pumped and occur in shocks along outflows, while CH$_{3}$OH masers
produced by radiative pumping and originate from circumstellar disks
or envelopes. Thus, it is not clear whether the correlation seen in
Fig. 3a arises from a physical connection between the far-infrared
luminosity and maser luminosity. Fig. 3b shows that there is also a
good correlation between methanol and water maser luminosities (R =
0.63, $p <$ 0.0001). Since there is no physical connection between
the two quantities, it is possible that the correlations are just a
distance squared effect, as suggested by Palla et al. (1991) for the
correlation between the water maser and the infrared luminosity. It
is also curious that the water masers in our sample (that are
associated with methanol masers) are an order of magnitude more
luminous than the ones associated with the methanol maser sample of
Szymczak et al. (2005). Some methanol masers detected in our survey
have low luminosities with the faintest source having a maser
luminosity of only $10^{-9}$ $L_{\odot}$. However, the infrared
luminosities of all sources range from $10^{3}$ to $10^{6}$
$L_{\odot}$, indicating that they are associated with massive star
formation.

\section{Summary}
A survey for 6.7 GHz CH$_{3}$OH masers was carried out toward 89
water masers and 10 new sources were detected, five of which are
located beyond the solar circle. A nil detection rate from low-mass
star forming regions enhances the conclusion that 6.7 GHz methanol
masers are only associated with massive star forming regions. There
is only one source maser with a peak flux density less than 1 Jy,
which could be due to the nature of the methanol maser luminosity
function. This paper presents only the results of the survey. In a
separate paper, we will report on the environment around the masers,
using ongoing observations of various molecules such as CO, HCO$^+$,
CN and NH$_3$. This will also elucidate on any differences between
faint masers and their bright counterparts.

\begin{acknowledgements}
We would like to thank the anonymous referee for many useful
suggestions and comments, which improved this paper. This research
is supported by NSFC under grants 10673024, 10733030, 10703010 and
10621303, and NBRPC (973 Program) under grant 2007CB815403.
\end{acknowledgements}

\newpage

\begin{table*}
\begin{flushleft}
\normalsize  Table 1: The target sample of H$_{2}$O maser sources.
\\[0.05mm]
\end{flushleft}
         \label{Tabiras}
      \[
           \begin{array} {lllllll}
             \hline \hline

\small

00211+6549             &  00259+5625^{\xi\dag}  &  00420+5530             &  00494+5617^{*}    &  01134+6429            &  W3 (1)              &\\
02219+6152^{*}         &  W3 (3)^{\xi}          &  02232+6138^{*}         &  02395+6244        &  02425+6851^{\dag}     &  02485+6902^{\dag}   &\\
03101+5821             &  03167+5848            &  03225+3034^{\dag}      &  03245+3002^{\dag} &  04579+4703            &  05137+3919^{**}     &\\
05168+3634^{\xi}       &  05274+3345^{*}        &  05302-0537^{\xi\dag}   &  05327-0457^{\xi}  &  KL IRC 2^{\xi}        &  05329-0508^{\xi\dag}&\\
05329-0512^{\dag}      &  05335+3609            &  05345+3556             &  05345+3157        &  05358+3543^{*}        &  05361+3539          &\\
05363+2454             &  05375+3540^{\xi}      &  NGC 2024^{\xi}         &  05445+0020^{\dag} &  05466+2316^{\xi\dag}  &  05553+1631          &\\
06001+3014^{\xi\dag}   &  MON R2^{\xi}          &  06053-0622^{*}         &  06055+2039^{*}    &  06058+2138^{*}        &  06061+2151^{*}      &\\
06067+2138^{\dag}      &  06099+1800^{*}        &  06117+1350^{*}         &  06127+1418        &  06291+0421            &  06306+0437^{\dag}   &\\
06437+0009             &  06446+0029^{**}       &  06501+0143             &  06579-0432^{\dag} &  06584-0852^{\xi}      &  07006-0654          &\\
18265+0028^{\dag}      &  18282-0951^{*}        &  18290-0924^{*}         &  18316-0602^{*}    &  18319-0802^{**}       &  G23.44-0.18^{*}     &\\
G23.01-0.41^{*}        &  OH24.7+0.2            &  18335-0713^{*}         &  18341-0727^{*}    &  18355-0650^{**}       &  18359-0334          &\\
18360-0537             &  18385-0512            &  18403-0440^{**}        &  18411-0338^{*}    &  18434-0242^{*}        &  G29.91-0.05^{**}    &\\
18449-0115^{*}         &  W43 (M2)^{\xi}        &  18450-0148^{\xi}       &  W43 (M3)^{*}      &  18455-0200            &  18461-0136          &\\
18469-0132             &  18469-0041            &  18470-0049^{*}         &  18479-0005^{**}   &  18487-0015^{*}        &  18507+0121^{*}      &\\
18507+0110^{*}         &  18517+0437^{*}        &  S76 W                  &  18537+0749        &  18538+0216^{\xi}      &  G35.05-0.52^{\xi}   &\\
18556+0136^{*}         &  18581+0409^{\xi}      &  18585+0407^{\xi}       &  18592+0108^{*}    &  18593+0408^{\xi}      &  19008+0530^{\xi}    &\\
19045+0813             &  19061+0652            &  19078+0901^{*}         &  W49 S^{\xi}       &  19088+0902            &  19092+0841^{*}      &\\
19095+0930^{*}         &  19110+1045^{*}        &  19181+1349             &  19207+1410        &  19209+1421            &  19213+1424^{*}      &\\
19213+1723             &  W51 M^{\xi}           &  19287+1816^{\dag}      &  19287+1845^{\xi\dag}  &  19303+1651^{*}        &  19363+2018          &\\
19374+2352             &  19388+2357^{*}        &  19410+2336^{*}         &  19433+2743^{\dag} &  19474+2637            &  19598+3324          &\\
20050+2720^{\dag}      &  20056+3350            &  20062+3550^{*}         &  20081+3122^{*}    &  20110+3321^{*}        &  20126+4104^{*}      &\\
20144+3726             &  20160+3911^{\xi}      &  20188+3928^{\xi}       &  20198+3716^{*}    &  ON 2 N                &  20215+3725^{\xi}    &\\
20227+4154^{\xi}       &  20231+3440^{\xi\dag}  &  20275+4001^{**}        &  20286+4105^{\xi}  &  W75 N^{*}             &  W75 S(1)^{\xi}      &\\
W75 OH^{\xi}           &  W75 S(3)^{\xi}        &  21007+4951^{\xi\dag}   &  21008+4700        &  21078+5211^{\xi}      &  21144+5430          &\\
21173+5450             &  21206+5145^{\xi}      &  21228+5332             &  G97.53+3.19^{\xi} &  21306+5540^{**}       &  21307+5049          &\\
21334+5039             &  21368+5502            &  21379+5106             &  21381+5000^{*}    &  21391+5026^{\xi}      &  21391+5802^{\dag}   &\\
21413+5442^{*}         &  21418+6552            &  BFS 11-B               &  21479+5510^{\xi}  &  21527+5727            &  21553+5908          &\\
21558+5907             &  22142+5206            &  22176+6303^{**}        &  22198+6336^{\dag} &  22199+6322^{\dag}     &  22305+5803       &\\
22475+5939             &  22517+6215^{\dag}     &  22543+6145^{*}         &  22570+5912^{\xi}  &  23004+5642            &  23004+5642       &\\
23030+5958^{\xi}       &  23116+6111^{*}        &  S158^{\xi}              &  23138+5945        &                                   &\\

             \hline \hline
         \end{array}
      \]
$^{*}$:\hspace{0.3cm}Previously known methanol masers.\\
$^{**}$:\hspace{0.2cm}New detected methanol masers.\\
$^{\xi}$:\hspace{0.3cm}Not observed due to constraints of observing time.\\
$^{\dag}$:\hspace{0.3cm}Infrared luminosities less than 10$^{3}$ $L_{\odot}$.\\

   \end{table*}

\begin{table*}
\begin{flushleft}
\normalsize  Table 2:  Details of the newly detected 6.7 GHz
CH$_{3}$OH masers. The first column lists the source name
(associated IRAS source or the galactic coordinates). The next two
columns give their J2000 equatorial coordinates. Col. (4) shows the
distances. Cols. (5) and (6) present the integrated flux density and
peak flux density. Cols. (7) and (8) show the radial velocity of peak emission
and the radial velocity range. Cols. (9) - (11) present the
infrared, CH$_{3}$OH and H$_{2}$O luminosities. The maser luminosities are
calculated assuming isotropic emission.
\\[0.05mm]
\end{flushleft}
         \label{Tabiras}
      \[
           \begin{array} {cccccccccccc}
             \hline \hline
      \noalign{\smallskip}
{\parbox[t]{20mm}{\centering Source \\ Name}}&
{\parbox[t]{17mm}{\centering R.A.(2000)\\
\mbox{$\mathrm{(^h\;\;\;^m\;\;\;^s)}$}}}&
{\parbox[t]{17mm}{\centering DEC(2000) \\
\mbox{$(\degr\;\;\;\arcmin\;\;\;\arcsec)$}}}&
{\parbox[t]{10mm}{\centering D \\(kpc)}}&
{\parbox[t]{15mm}{\centering $S_i$\\(Jy~km~s$^{-1}$)}}&
{\parbox[t]{10mm}{\centering $S_p$\\(Jy)}}&
{\parbox[t]{10mm}{\centering $v_p$\\(km~s$^{-1}$)}}&
{\parbox[t]{12mm}{\centering $\Delta v$
\\(km~s$^{-1}$)}}&{\parbox[t]{10mm}{\centering
$L_{IR}$\\($L_{\odot}$)}}& {\parbox[t]{12mm}{\centering
$L_{CH_{3}OH}$\\($L_{\odot}$)}}& {\parbox[t]{10mm}{\centering
$L_{H_{2}O}$\\($L_{\odot}$)}}

\\
       \noalign{\smallskip}
\hline
      \noalign{\smallskip}
\small

 05137+3919  & 05\ 17\ 13.3 & $+$39\ 22\ 14 &  12.0^{1}& 0.7 & 1.5 & $-$16.2 & $-$21.1,$-$16.0  & 6.1(4)        & 7.0($-$7)         & 2.2($-$4) \\
 06446+0029  & 06\ 47\ 12.9 & $+$00\ 26\ 07 &  6.0^{1} & 2.8 & 1.6 &    48.6 &    42.7,50.0     & 2.3(4)        & 7.0($-$7)         & 1.5($-$4) \\
 18319-0802  & 18\ 34\ 38.2 & $-$07\ 59\ 35 &  2.8^{\ast}  & 2.2 & 2.1 &    35.7 &    31.0,45.2     & 7.2(3)        & 1.2($-$7)         & 4.5($-$6) \\
 18355-0650  & 18\ 38\ 14.3 & $-$06\ 47\ 47 &  5.7^{4} & 4.2 & 3.9 &    58.2 &    56.6,62.1     & 7.4(5)        & 9.4($-$7)         & 7.1($-$6) \\
 18403-0440  & 18\ 43\ 01.1 & $-$04\ 36\ 41 &  3.3^{4} & 0.3 & 1.0 &    20.1 &    19.8,20.4     & 4.1(3)        & 2.3($-$8)         & 3.2($-$6) \\
 G29.91-0.05 & 18\ 46\ 05.9 & $-$02\ 42\ 27 &  7.1^{\ast}  & 7.8 & 2.9 &   104.3 &    93.4,105.3    &  ---          & 2.7($-$6)         & 3.3($-$5) \\
 18479-0005  & 18\ 50\ 31.1 & $-$00\ 01\ 54 &  13.0^{2}& 1.3 & 1.5 &    27.4 &    24.9,28.3     & 1.6(6)        & 1.5($-$6)         & 9.0($-$5) \\
 20275+4001  & 20\ 29\ 24.9 & $+$40\ 11\ 21 &  2.0^{3} & 1.7 & 4.0 &  $-$6.9 &  $-$7.4,$-$6.5   & 6.7(4)        & 4.7($-$8)         & 1.5($-$5) \\
 21306+5540  & 21\ 32\ 13.0 & $+$55\ 52\ 56 &  8.6^{1} & 1.5 & 1.1 & $-$69.9 &  $-$73.7,$-$68.0 & 1.4(5)        & 7.8($-$7)         & 9.9($-$4) \\
 22176+6303  & 22\ 19\ 18.3 & $+$63\ 18\ 48 &  0.9^{6} & 0.2 & 0.5 &  $-$2.0 &  $-$2.3,$-$1.1   & 2.4(4)        & 1.1($-$9)         & 4.8($-$7) \\

\noalign{\smallskip} \hline

         \end{array}
      \]

 References for the distances: \\
 1 \  Wouterloot et al. 1993. \ 2 \  Palagi et al 1993. \ 3 \
 Anglada et al. 1996. \ 4 \ van der Tak et al. 1999. \ 5 \  Wu et al. 2006. \ 6 \ Crampton \& Fisher 1974 \\
$^{\ast}$ Heliocentric near kinematic distances.

\end{table*}
\begin{figure*}
\begin{tabular}{cccc}
\includegraphics[width=4cm,angle=-90]{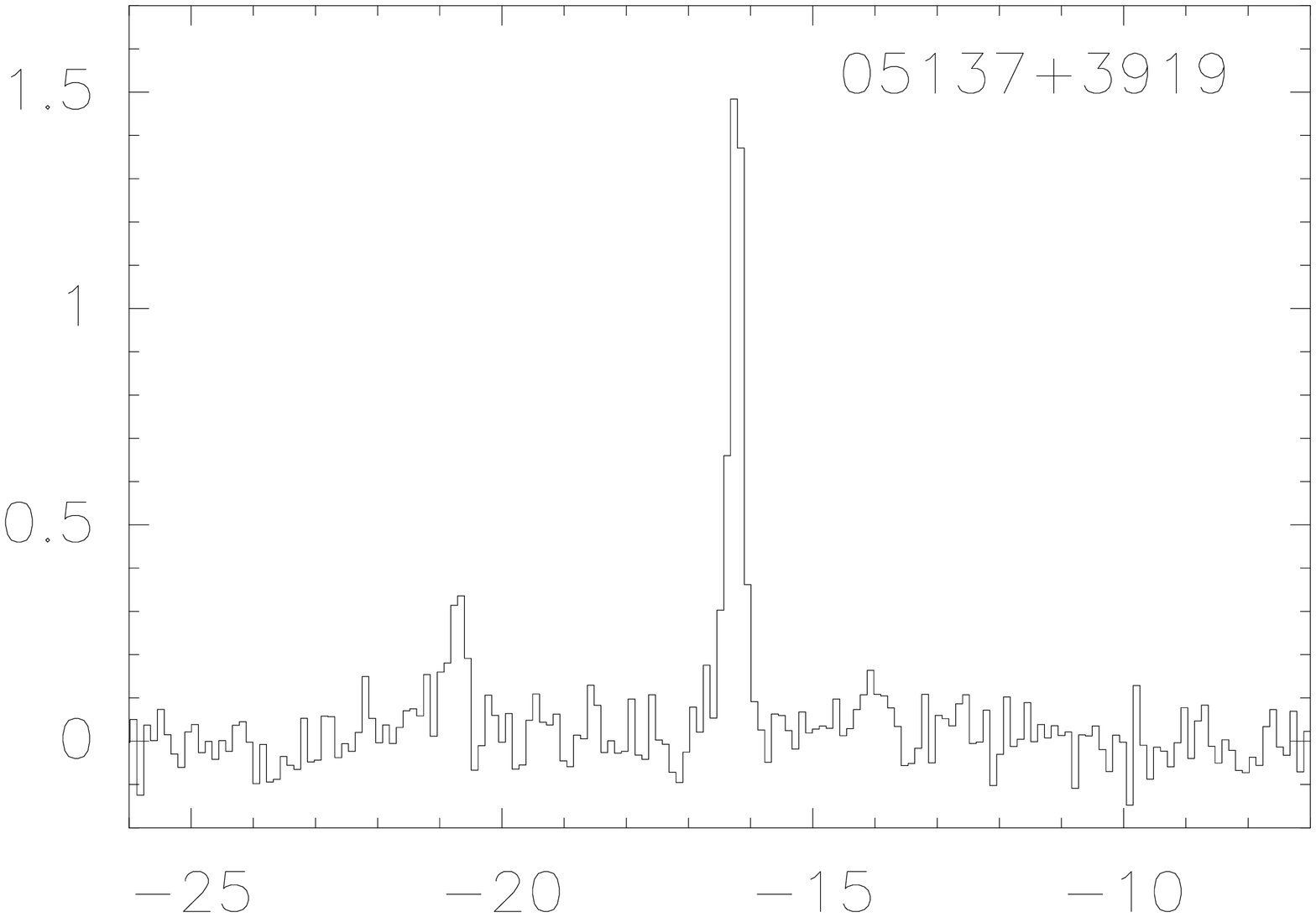} &
\includegraphics[width=4cm,angle=-90]{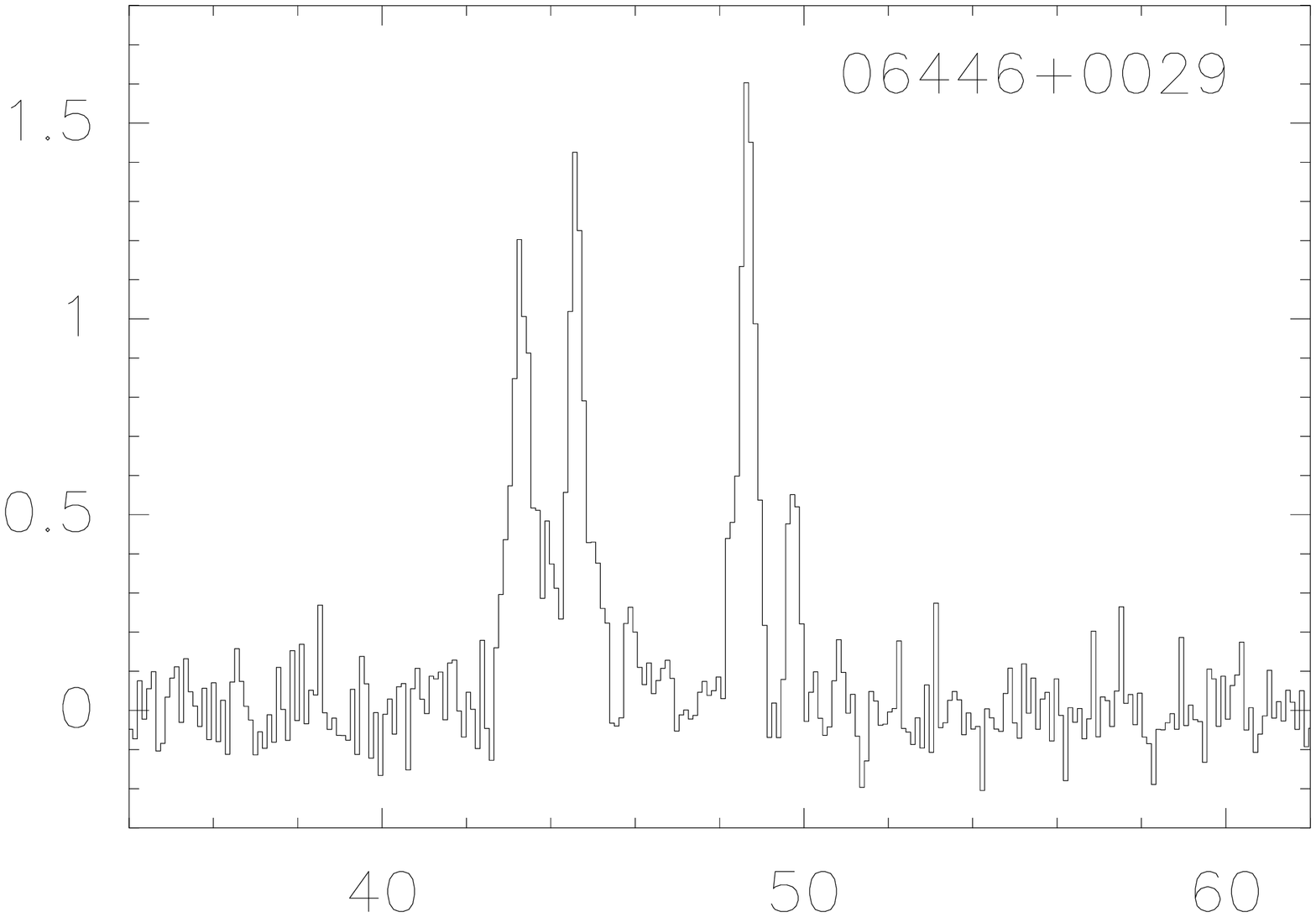} &
\includegraphics[width=4cm,angle=-90]{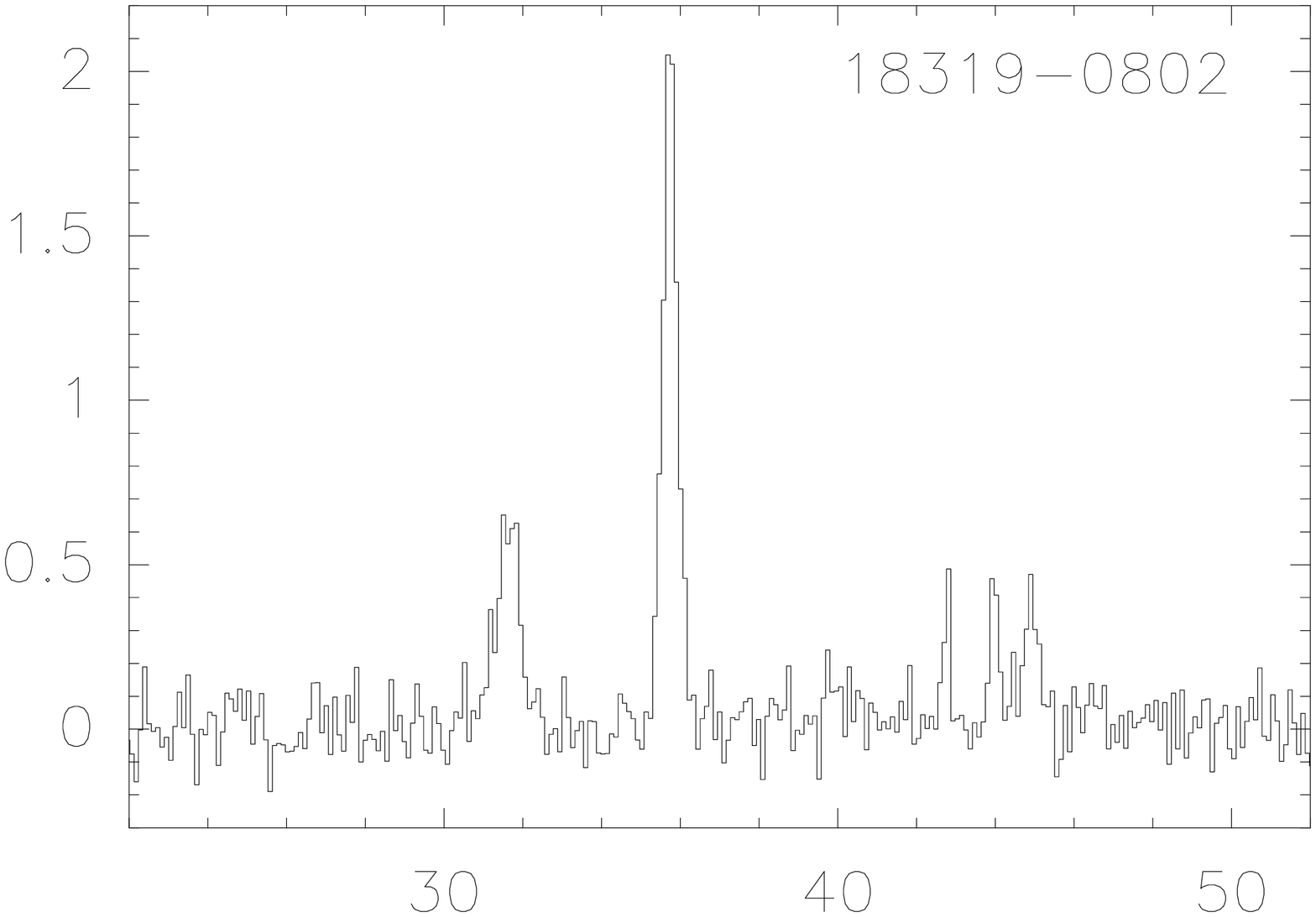} &
\\
\includegraphics[width=4cm,angle=-90]{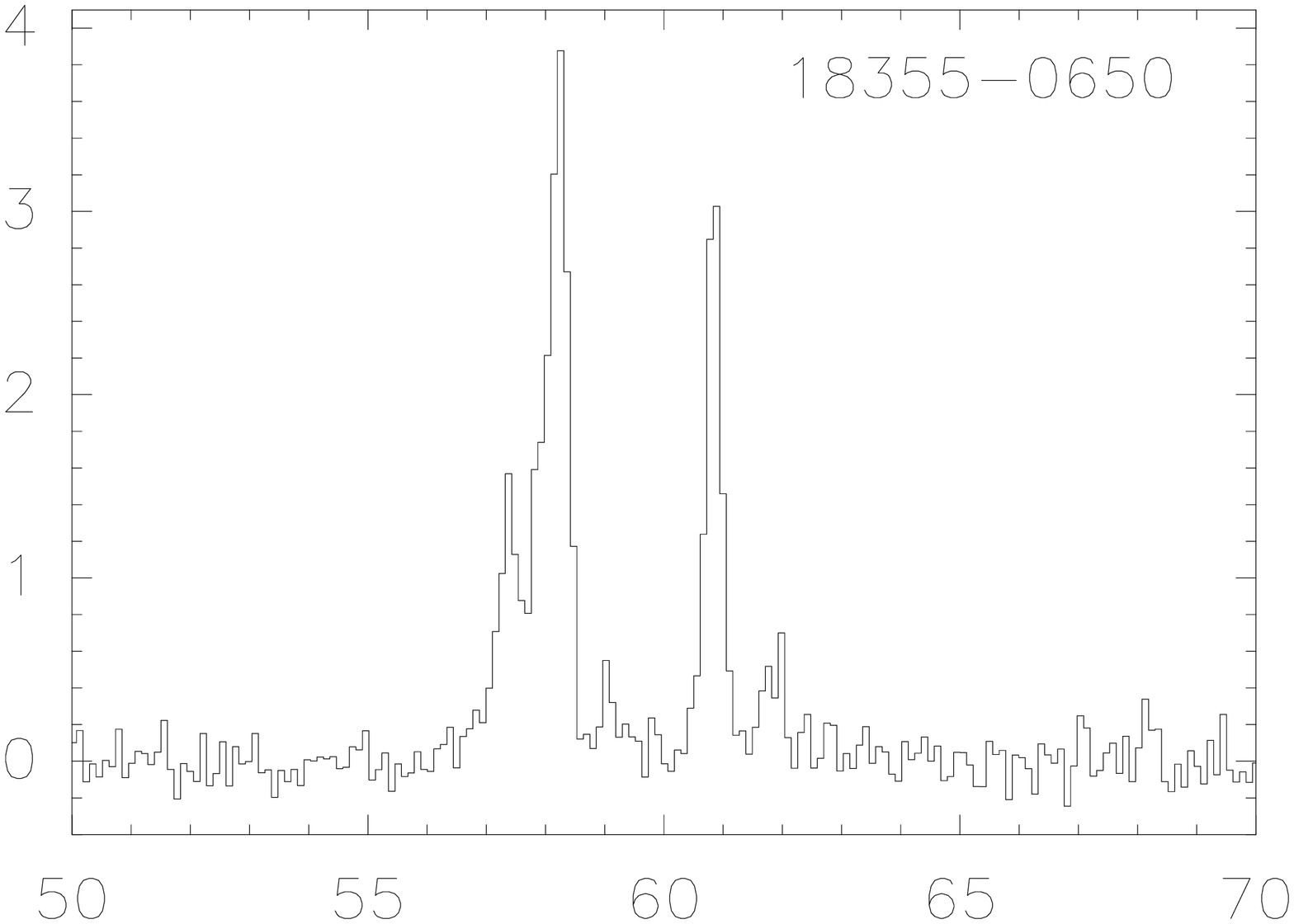} &
\includegraphics[width=4cm,angle=-90]{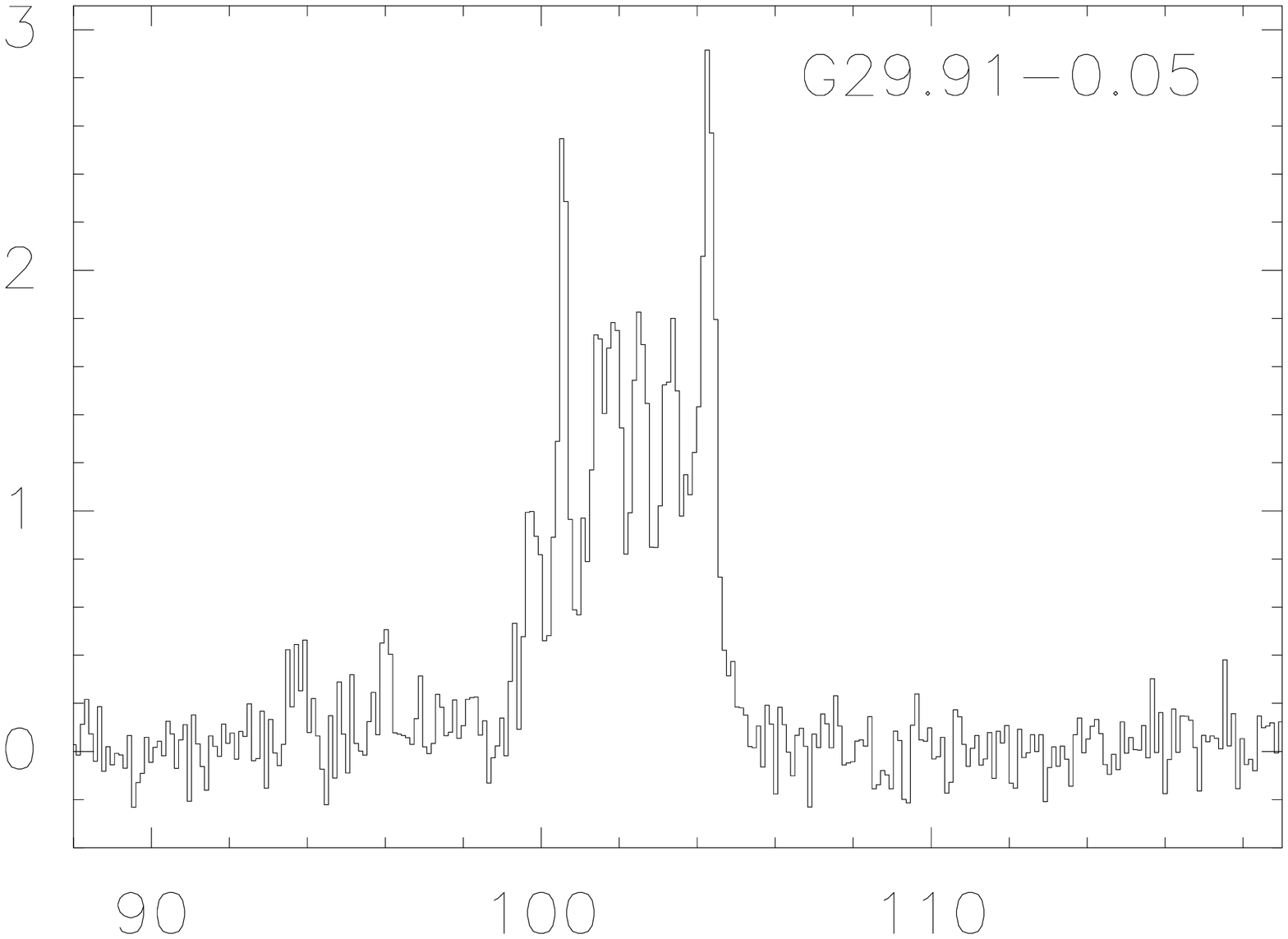} &
\includegraphics[width=4cm,angle=-90]{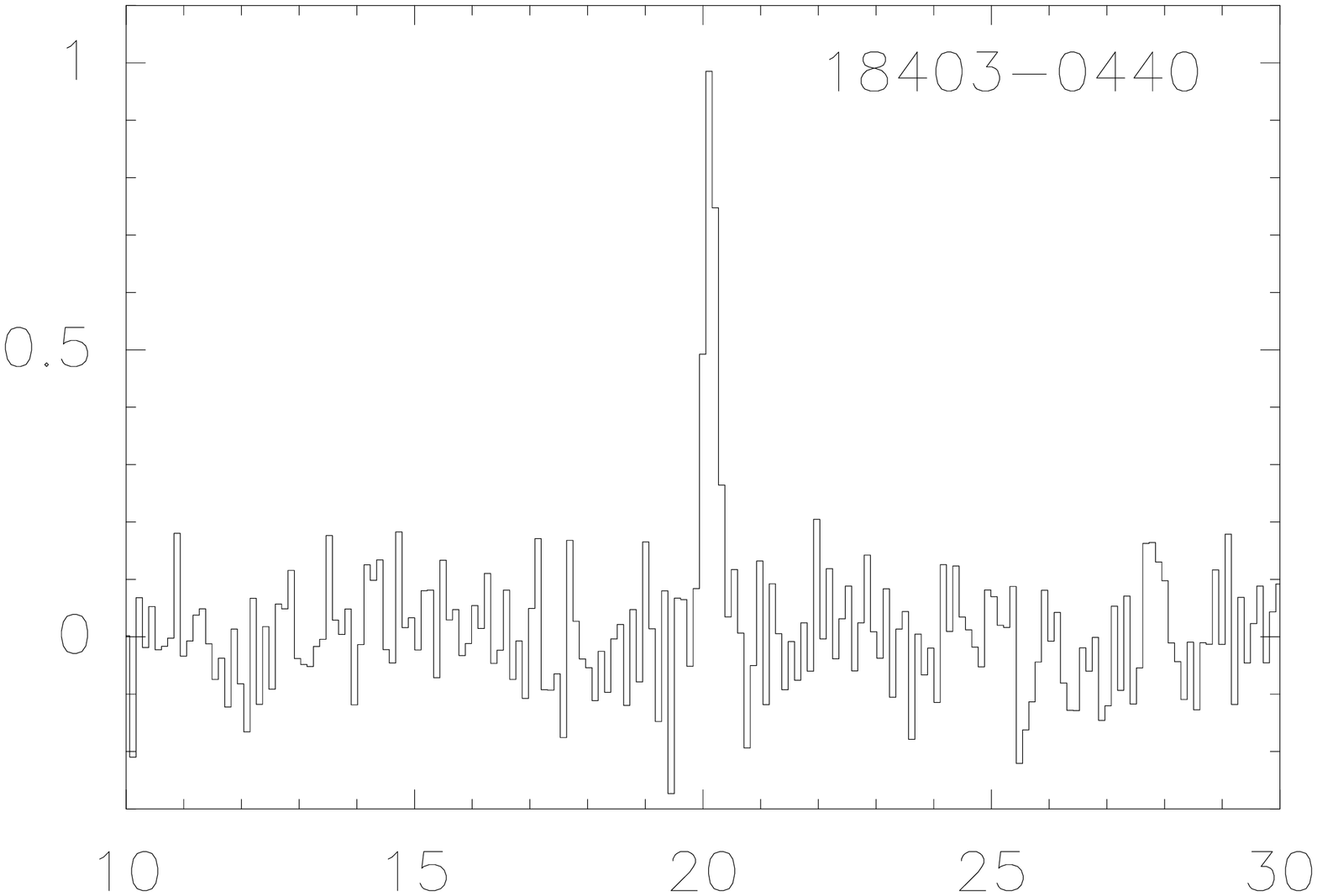} &
\\
\includegraphics[width=4cm,angle=-90]{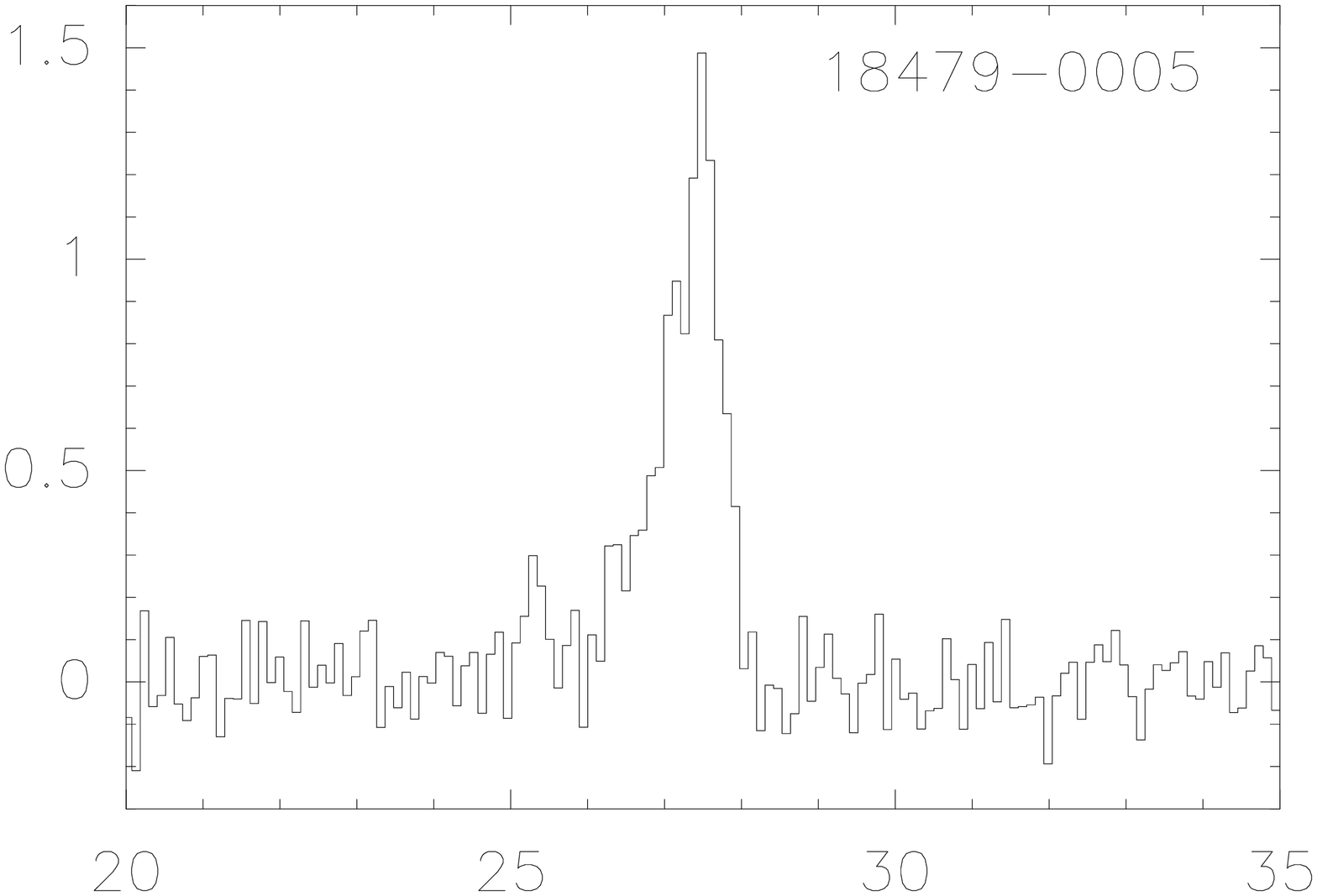} &
\includegraphics[width=4cm,angle=-90]{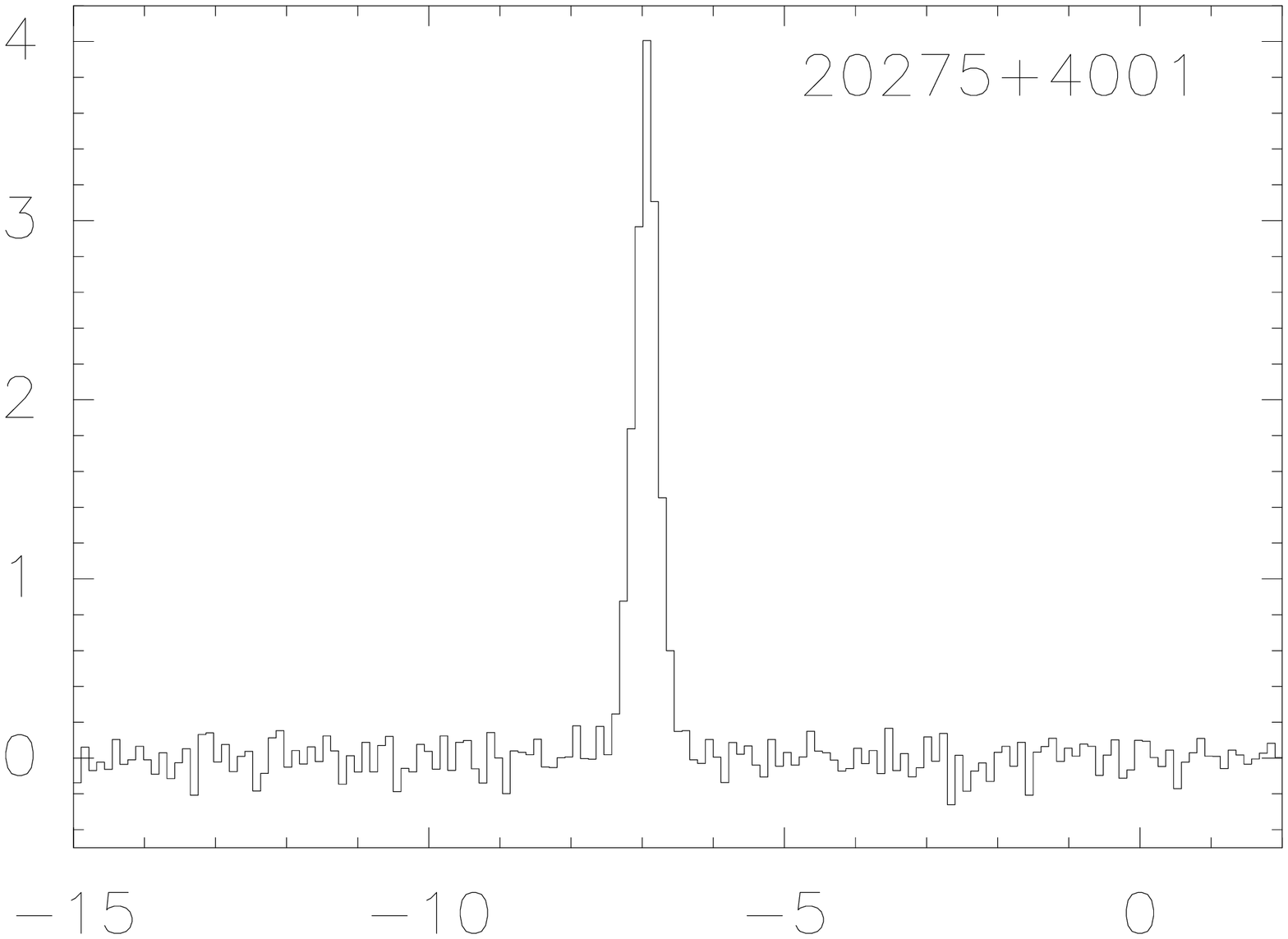} &
\includegraphics[width=4cm,angle=-90]{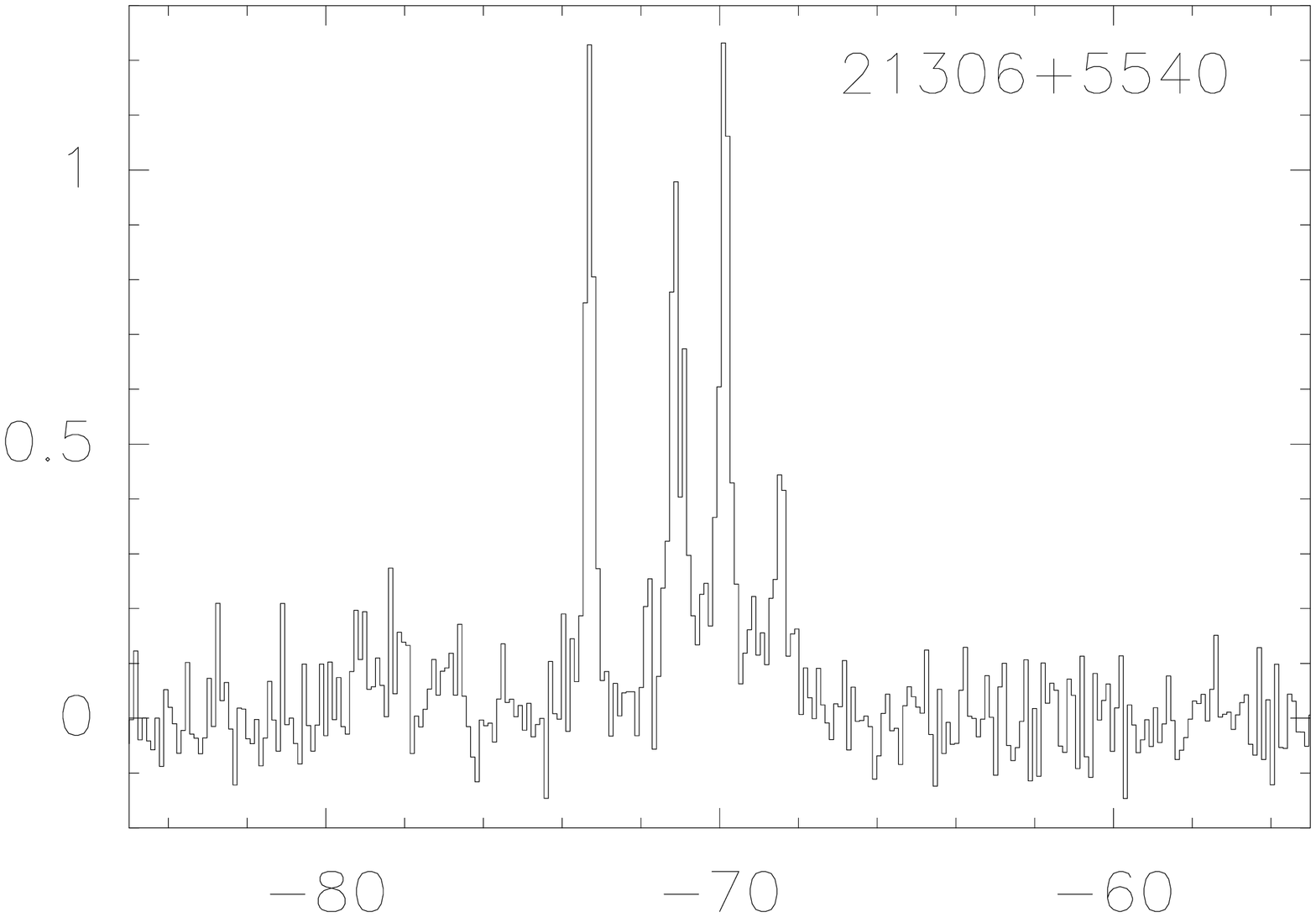} &
\\
\includegraphics[width=4cm,angle=-90]{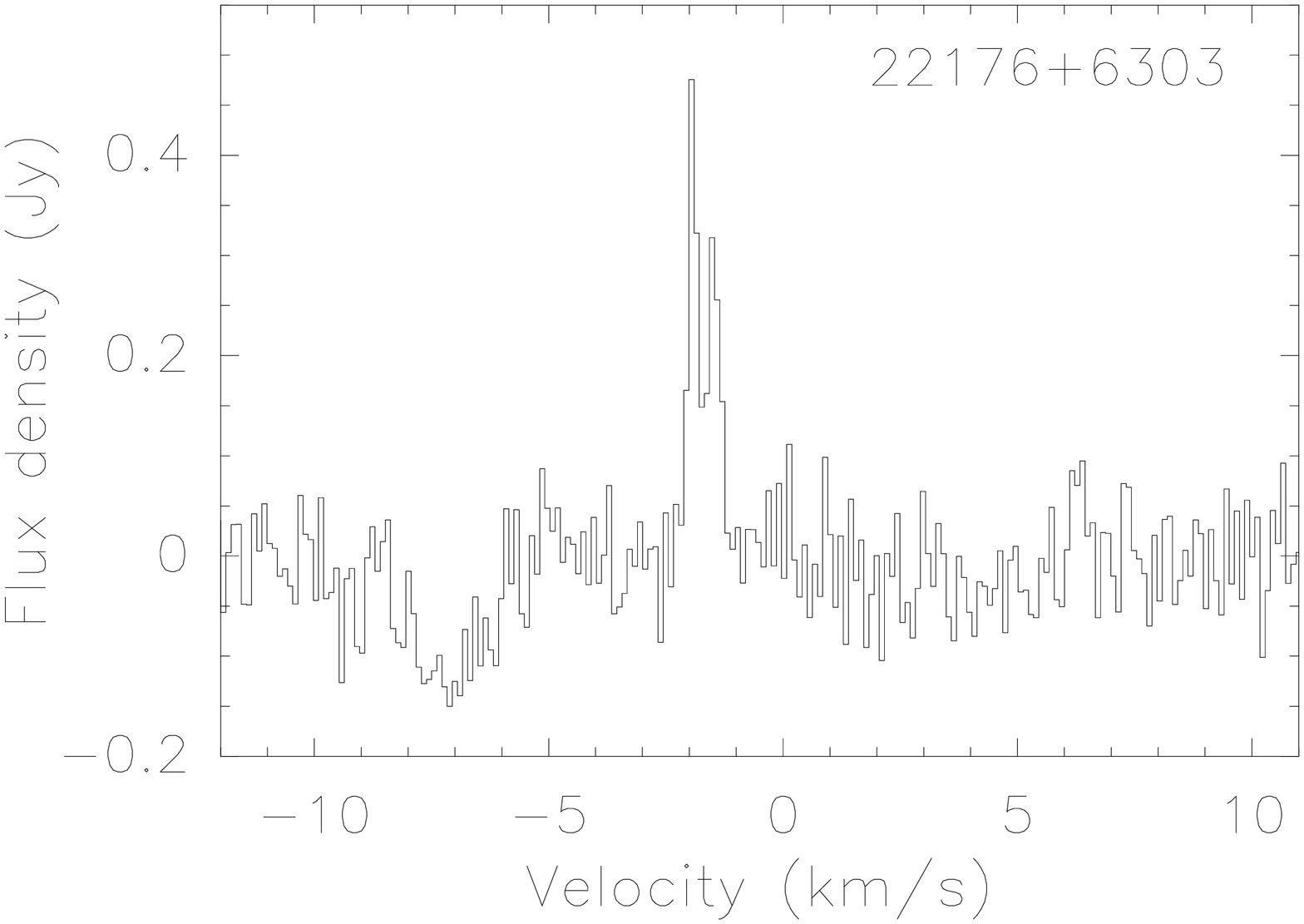} &
\\
\end{tabular}
\caption{Spectra of the new 6.7-GHz CH$_{3}$OH maser detections. The spectral
resolution is 0.11 km s$^{-1}$.}
\end{figure*}

\clearpage
\begin{figure}
\includegraphics[]{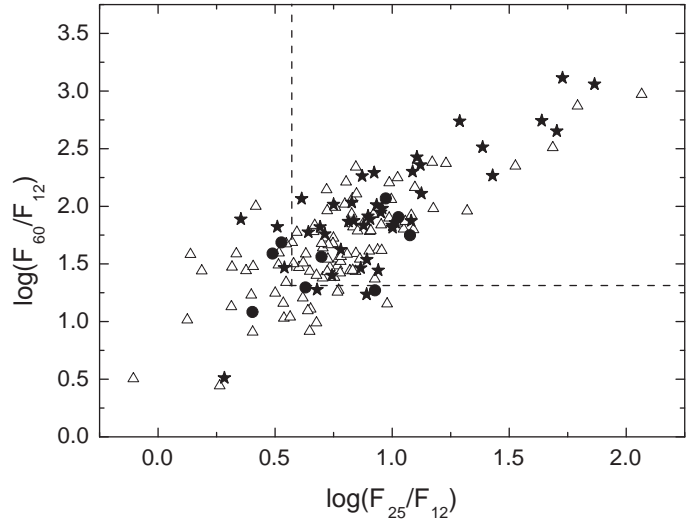}
\caption{IRAS color-color diagram for 154 sources. The box in the
upper right corner delineates the WC criteria for UCHII regions.
Sources with no methanol maser detections are shown in open
triangles. Previous and new detections of methanol masers are shown
in stars and circles respectively.} \label{}
\end{figure}
\begin{figure*}
\begin{tabular}{ccc}
\includegraphics[width=7cm]{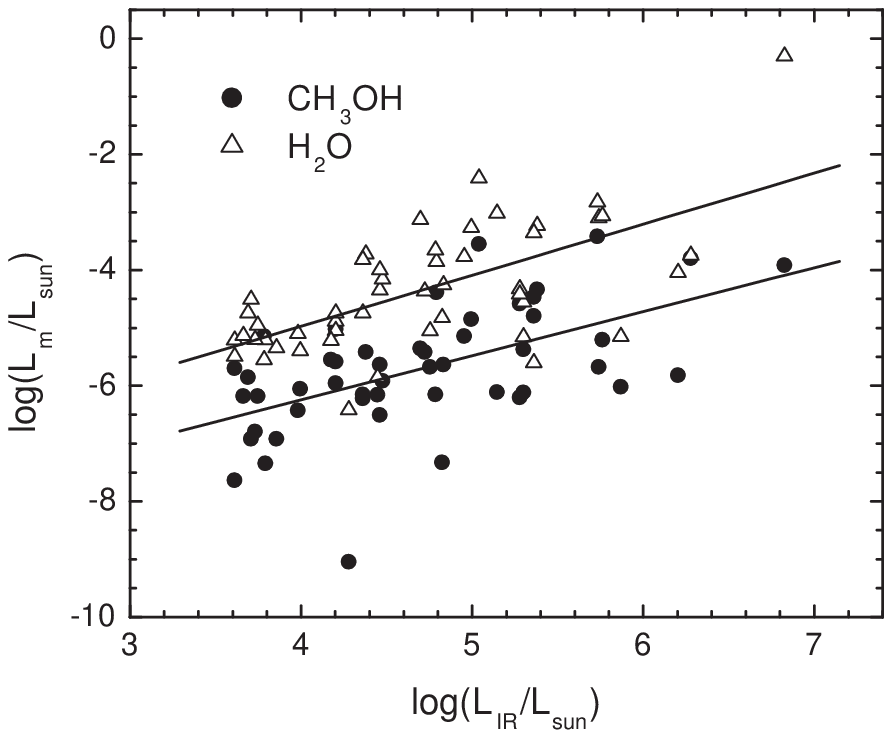} &
\includegraphics[width=7cm]{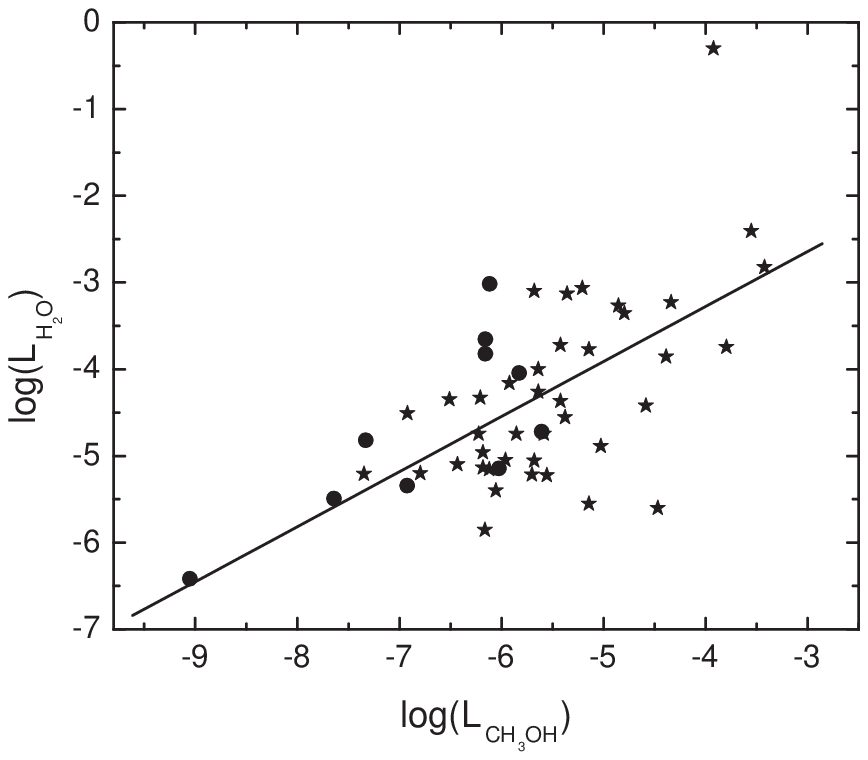} &
\end{tabular}
\caption{Left panel -- (a) Methanol and water maser luminosities versus the infrared
luminosity calculated from the IRAS data. The circles and triangles
show the methanol and water masers respectively. Water masers
without any methanol maser associations are excluded. Right panel -- (b) Water maser
versus methanol maser luminosity. Previous and new detections of
methanol masers are shown in stars and circles respectively.}
\end{figure*}
\clearpage
\section{Online Material}
\begin{table*}
\begin{flushleft}
\normalsize  Table 3: H$_{2}$O maser sources that have non
detections for 6.7 GHz CH$_{3}$OH maser emission. The columns show
the source name, J2000 coordinates, central velocity for the
spectrometer, $v_c$, velocity range covered by the spectrometer,
$v_{min}$ and $v_{max}$, the velocity resolution, $\delta(v)$, and
the rms noise in the spectrum ($1\sigma$ limit on the maser flux
density).
\\[0.05mm]
\end{flushleft}
         \label{Tabiras}
      \[
           \begin{array} {cccccccc}
             \hline \hline
      \noalign{\smallskip}
{\parbox[t]{19mm}{\centering Source \\ Name}}&
{\parbox[t]{17mm}{\centering R.A.(2000)\\
\mbox{$\mathrm{(^h\;\;\;^m\;\;\;^s)}$}}}&
{\parbox[t]{17mm}{\centering DEC(2000) \\
\mbox{$(\degr\;\;\;\arcmin\;\;\;\arcsec)$}}}&
{\parbox[t]{14mm}{\centering $v_c$\\(km~s$^{-1}$)}}&
{\parbox[t]{18mm}{\centering $v_{min}$\\(km~s$^{-1}$)}}&
{\parbox[t]{18mm}{\centering $v_{max}$\\(km~s$^{-1}$)}}&
{\parbox[t]{14mm}{\centering $\delta(v)$\\(km~s$^{-1}$)}}&
{\parbox[t]{18mm}{\centering RMS\\ (Jy)}}
\\
       \noalign{\smallskip}
\hline
      \noalign{\smallskip}
\small

00211+6549        & 00\ 23\ 58.1   & $+$66\ 06\ 03   & $-$71.1 & $-$296  & 154 & 0.11   & 0.14 \\
00420+5530        & 00\ 44\ 57.5   & $+$55\ 47\ 18   & $-$47.8 & $-$273  & 177 & 0.11   & 0.14 \\
01134+6429        & 01\ 16\ 47.2   & $+$64\ 45\ 39   & $-$46.9 & $-$272  & 178 & 0.11   & 0.10 \\
W3 (1)            & 02\ 25\ 28.2   & $+$62\ 06\ 58   & $-$40.0 & $-$265  & 185 & 0.11   & 0.10 \\
02395+6244        & 02\ 43\ 29.1   & $+$62\ 57\ 00   & $-$67.1 & $-$292  & 158 & 0.11   & 0.10 \\
02425+6851        & 02\ 47\ 00.2   & $+$69\ 04\ 11   & $-$10.4 & $-$461  & 440 & 0.22   & 0.05 \\
02485+6902        & 02\ 53\ 07.2   & $+$69\ 14\ 36   & $-$10.4 & $-$461  & 440 & 0.22   & 0.05 \\
03101+5821        & 03\ 14\ 04.7   & $+$58\ 33\ 08   & $-$38.5 & $-$489  & 412 & 0.22   & 0.05 \\
03167+5848        & 03\ 20\ 44.3   & $+$58\ 59\ 33   & $-$55.6 & $-$506  & 395 & 0.22   & 0.05 \\
03225+3034        & 03\ 25\ 35.5   & $+$30\ 45\ 21   & 2.6     & $-$223  & 228 & 0.11   & 0.19 \\
03245+3002        & 03\ 27\ 39.0   & $+$30\ 13\ 00   & 4.4     & $-$446  & 455 & 0.22   & 0.05 \\
04579+4703        & 05\ 01\ 39.7   & $+$47\ 07\ 23   & $-$17.8 & $-$243  & 207 & 0.11   & 0.10 \\
05329-0512        & 05\ 35\ 27.5   & $-$05\ 09\ 37   & 7.5     & $-$218  & 233 & 0.11   & 0.19 \\
05335+3609        & 05\ 36\ 52.5   & $+$36\ 10\ 49   & $-$17.3 & $-$468  & 433 & 0.22   & 0.05 \\
05345+3556        & 05\ 37\ 57.8   & $+$35\ 58\ 41   & $-$20.7 & $-$471  & 430 & 0.22   & 0.05 \\
05345+3157        & 05\ 37\ 53.1   & $+$31\ 59\ 35   & $-$20.0 & $-$245  & 205 & 0.11   & 0.19 \\
05361+3539        & 05\ 39\ 27.7   & $+$35\ 40\ 43   & $-$18.0 & $-$469  & 433 & 0.22   & 0.05 \\
05363+2454        & 05\ 39\ 28.2   & $+$24\ 56\ 32   & 13.9    & $-$437  & 464 & 0.22   & 0.05 \\
05445+0020        & 05\ 47\ 05.4   & $+$00\ 21\ 50   & 15.9    & $-$209  & 241 & 0.11   & 0.19 \\
05553+1631        & 05\ 58\ 13.9   & $+$16\ 32\ 00   & $-$7.6  & $-$233  & 218 & 0.11   & 0.14 \\
06067+2138        & 06\ 09\ 48.0   & $+$21\ 38\ 11   & 3.2     & $-$447  & 454 & 0.22   & 0.05 \\
06127+1418        & 06\ 15\ 34.6   & $+$14\ 17\ 10   & 11.3    & $-$214  & 237 & 0.11   & 0.19 \\
06291+0421        & 06\ 31\ 48.1   & $+$04\ 19\ 31   & 9.6     & $-$216  & 235 & 0.11   & 0.19 \\
06306+0437        & 06\ 33\ 16.4   & $+$04\ 34\ 57   & 10.1    & $-$440  & 461 & 0.22   & 0.05 \\
06437+0009        & 06\ 46\ 15.6   & $+$00\ 06\ 19   & 33.6    & $-$192  & 259 & 0.11   & 0.19 \\
06501+0143        & 06\ 52\ 45.6   & $+$01\ 40\ 15   & 52.0    & $-$399  & 503 & 0.22   & 0.05 \\
06579-0432        & 07\ 00\ 23.1   & $-$04\ 36\ 38   & 26.3    & $-$424  & 477 & 0.22   & 0.05 \\
07006-0654        & 07\ 03\ 05.1   & $-$06\ 58\ 28   & 29.1    & $-$421  & 480 & 0.22   & 0.05 \\
18265+0028        & 18\ 29\ 05.8   & $+$00\ 30\ 36   & 5.3     & $-$220  & 231 & 0.11   & 0.19 \\
OH24.7+0.2        & 18\ 35\ 29.9   & $-$07\ 13\ 10   & 25.9    & $-$199  & 251 & 0.11   & 0.14 \\
18359-0334        & 18\ 38\ 34.3   & $-$03\ 32\ 06   & $-$45.5 & $-$271  & 180 & 0.11   & 0.14 \\
18360-0537        & 18\ 38\ 41.6   & $-$05\ 35\ 06   & 104.4   & $-$121  & 330 & 0.11   & 0.19 \\

\noalign{\smallskip} \hline

         \end{array}
      \]

   \end{table*}

\begin{table*}
\begin{flushleft}
\normalsize  Table 1-continued
\\[0.05mm]
\end{flushleft}
         \label{Tabiras}
      \[
           \begin{array} {cccccccc}
             \hline \hline
      \noalign{\smallskip}
{\parbox[t]{19mm}{\centering Source \\ Name}}&
{\parbox[t]{17mm}{\centering R.A.(2000)\\
\mbox{$\mathrm{(^h\;\;\;^m\;\;\;^s)}$}}}&
{\parbox[t]{17mm}{\centering DEC(2000) \\
\mbox{$(\degr\;\;\;\arcmin\;\;\;\arcsec)$}}}&
{\parbox[t]{14mm}{\centering $v_c$\\(km~s$^{-1}$)}}&
{\parbox[t]{18mm}{\centering $v_{min}$\\(km~s$^{-1}$)}}&
{\parbox[t]{18mm}{\centering $v_{max}$\\(km~s$^{-1}$)}}&
{\parbox[t]{14mm}{\centering $\delta(v)$\\(km~s$^{-1}$)}}&
{\parbox[t]{18mm}{\centering RMS\\ (Jy)}}
\\
       \noalign{\smallskip}
\hline
      \noalign{\smallskip}
\small
18385-0512        & 18\ 41\ 12.0   & $-$05\ 09\ 07   & 18.0    & $-$207  & 243 & 0.11   & 0.24 \\
18455-0200        & 18\ 48\ 08.8   & $-$01\ 56\ 54   & 104.3   & $-$130  & 320 & 0.11   & 0.19 \\
18461-0136        & 18\ 48\ 45.2   & $-$01\ 33\ 12   & 18.6    & $-$207  & 244 & 0.11   & 0.19 \\
18469-0132        & 18\ 49\ 32.8   & $-$01\ 29\ 04   & 93.6    & $-$132  & 319 & 0.11   & 0.24 \\
18469-0041        & 18\ 49\ 32.2   & $-$00\ 38\ 01   & 92.5    & $-$133  & 318 & 0.11   & 0.19 \\
S76 W             & 18\ 55\ 58.7   & $+$07\ 53\ 42   & $-$3.6  & $-$229  & 222 & 0.11   & 0.19 \\
18537+0749        & 18\ 56\ 11.8   & $+$07\ 53\ 24   & 31.7    & $-$194  & 257 & 0.11   & 0.19 \\
19045+0813        & 19\ 06\ 59.7   & $+$08\ 18\ 42   & 16.9    & $-$208  & 242 & 0.11   & 0.19 \\
19061+0652        & 19\ 08\ 36.8   & $+$06\ 57\ 02   & 71.9    & $-$153  & 297 & 0.11   & 0.19 \\
19088+0902        & 19\ 11\ 15.8   & $+$09\ 07\ 27   & $-$5.1  & $-$230  & 220 & 0.11   & 0.19 \\
19181+1349        & 19\ 20\ 31.2   & $+$13\ 55\ 24   & 34.1    & $-$191  & 259 & 0.11   & 0.19 \\
19207+1410        & 19\ 23\ 01.2   & $+$14\ 16\ 40   & 68.6    & $-$157  & 294 & 0.11   & 0.29 \\
19209+1421        & 19\ 23\ 10.9   & $+$14\ 26\ 38   & 52.1      & $-$173  & 277 & 0.11 & 0.24 \\
19213+1723        & 19\ 23\ 37.0   & $+$17\ 28\ 59   & $-$26.8   & $-$252  & 198 & 0.11 & 0.14 \\
19287+1816        & 19\ 30\ 58.3   & $+$18\ 22\ 28   & 24.5      & $-$201  & 250 & 0.11 & 0.19 \\
19363+2018        & 19\ 38\ 31.6   & $+$20\ 25\ 21   & 36.0      & $-$189  & 261 & 0.11 & 0.19 \\
19374+2352        & 19\ 39\ 33.2   & $+$23\ 59\ 55   & 39.4      & $-$186  & 265 & 0.11 & 0.19 \\
19433+2743        & 19\ 45\ 20.8   & $+$27\ 50\ 51   & 19.9      & $-$205  & 245 & 0.11 & 0.19 \\
19474+2637        & 19\ 49\ 32.5   & $+$26\ 45\ 14   & 20.5      & $-$205  & 246 & 0.11 & 0.19 \\
19598+3324        & 20\ 01\ 45.5   & $+$33\ 32\ 41   & $-$19.4   & $-$245  & 206 & 0.11 & 0.24 \\
20050+2720        & 20\ 07\ 06.7   & $+$27\ 28\ 53   & 0.9       & $-$224  & 226 & 0.11 & 0.19 \\
20056+3350        & 20\ 07\ 31.5   & $+$33\ 59\ 39   & 10.4      & $-$215  & 236 & 0.11 & 0.19 \\
20144+3726        & 20\ 16\ 15.8   & $+$37\ 35\ 40   & $-$56.7   & $-$507  & 394 & 0.22 & 0.05 \\
ON 2 N            & 20\ 21\ 43.9   & $+$37\ 26\ 39   & 2.4       & $-$448  & 453 & 0.22 & 0.05 \\
21008+4700        & 21\ 02\ 32.3   & $+$47\ 12\ 36   & $-$49.6   & $-$500  & 401 & 0.22 & 0.05 \\
21144+5430        & 21\ 15\ 55.8   & $+$54\ 43\ 31   & $-$83.2   & $-$534  & 367 & 0.22 & 0.05 \\
21173+5450        & 21\ 18\ 53.1   & $+$55\ 03\ 19   & $-$82.8   & $-$533  & 368 & 0.22 & 0.05 \\
21228+5332        & 21\ 24\ 29.0   & $+$53\ 45\ 34   & $-$97.4   & $-$548  & 353 & 0.22 & 0.05 \\
21307+5049        & 21\ 32\ 31.4   & $+$51\ 02\ 23   & $-$46.7   & $-$272  & 179 & 0.11 & 0.10 \\
21334+5039        & 21\ 35\ 09.1   & $+$50\ 53\ 09   & $-$43.9   & $-$269  & 181 & 0.11 & 0.19 \\
21368+5502        & 21\ 38\ 25.8   & $+$55\ 16\ 33   & $-$83.2   & $-$308  & 142 & 0.11 & 0.19 \\
21379+5106        & 21\ 39\ 40.5   & $+$51\ 20\ 32   & $-$41.6   & $-$267  & 184 & 0.11 & 0.19 \\
21391+5802        & 21\ 40\ 42.3   & $+$58\ 16\ 10   & $-$5.0    & $-$230  & 220 & 0.11 & 0.10 \\
21418+6552        & 21\ 43\ 04.0   & $+$66\ 05\ 57   & $-$31.2   & $-$256  & 194 & 0.11 & 0.10 \\
BFS 11-B          & 21\ 43\ 06.0   & $+$66\ 06\ 57   & $-$14.9   & $-$240  & 210 & 0.11 & 0.05 \\
21527+5727        & 21\ 54\ 21.5   & $+$57\ 41\ 14   & $-$70.3   & $-$296  & 155 & 0.11 & 0.10 \\
21553+5908        & 21\ 56\ 59.3   & $+$59\ 22\ 39   & $-$87.8   & $-$538  & 363 & 0.22 & 0.05 \\
21558+5907        & 21\ 57\ 24.5   & $+$59\ 21\ 56   & $-$89.6   & $-$315  & 136 & 0.11 & 0.05 \\
22142+5206        & 22\ 16\ 10.4   & $+$52\ 21\ 25   & $-$37.4   & $-$488  & 413 & 0.22 & 0.05 \\
22198+6336        & 22\ 21\ 27.5   & $+$63\ 51\ 46   & $-$21.3   & $-$247  & 204 & 0.11 & 0.10 \\
22199+6322        & 22\ 21\ 33.2   & $+$63\ 37\ 22   & $-$9.0    & $-$460  & 442 & 0.22 & 0.05 \\
22305+5803        & 22\ 32\ 24.2   & $+$58\ 18\ 58   & $-$53.0   & $-$278  & 172 & 0.11 & 0.19 \\
22475+5939        & 22\ 49\ 29.5   & $+$59\ 55\ 37   & $-$54.5   & $-$280  & 171 & 0.11 & 0.10 \\
22517+6215        & 22\ 53\ 43.4   & $+$62\ 31\ 46   & $-$9.0    & $-$234  & 216 & 0.11 & 0.19 \\
23004+5642        & 23\ 02\ 34.9   & $+$56\ 58\ 55   & $-$53.5   & $-$504  & 397 & 0.22 & 0.05 \\
23004+5642        & 23\ 02\ 32.1   & $+$56\ 57\ 51   & $-$53.5   & $-$279  & 172 & 0.11 & 0.19\\
23138+5945        & 23\ 16\ 04.4   & $+$60\ 01\ 34   & $-$44.5   & $-$270  & 181 & 0.11 & 0.19 \\

\noalign{\smallskip} \hline

         \end{array}
      \]

 \end{table*}


\begin{thebibliography}{}
  \bibitem{} Anglada, G., Estalella, R., and Pastor, J. et al., 1996, \apj, 463, 205
  \bibitem{} Becker, Robert H., White, Richard L., Helfand, David J., \& Zoonematkermani, S., 1994, \apjs, 91, 347
  \bibitem{} Beuther, H., Walsh, A., Schilke, P., Sridharan, T. K., Menten, K. M., \& Wyrowski, F., 2002, \aap, 390, 289
  \bibitem{} Brand, J., Cesaroni, R., Caselli, P., et al. 1994, \aaps, 103, 541
  \bibitem{} Breen, S. L., Ellingsen, S. P., \& Johnston-Hollitt, M. et al., 2007, \mnras, 377, 491
  \bibitem{} Breckenridge, S. M., \& Kukolich, S. G. 1995, \apj, 438, 504
  \bibitem{} Casoli F., Dupraz C., Gerin M., Combes F., Boulanger F. 1986, A\&A 169, 281
  \bibitem{} Caswell, J. L., Vaile, R. A., Ellingsen, S. P., Whiteoak, J. B., \& Norris, R. P, 1995, \mnras, 272, 96
  \bibitem{} Codella, C., \& Moscadelli, L., 2000, \aap, 362, 723
  \bibitem{} Comoretto, G., Palagi, F., Cesaroni, R., et al. 1990, \aaps, 84, 179
  \bibitem{} Crampton, D., \& Fisher, W. A., 1974, Pub. Dom. Astrophys. Obs., 14, 283
  \bibitem{} Ellingsen, S. P. 2006, \apj, 638, 241
  \bibitem{} Ellingsen, S. P. 2007, \mnras, 377, 571
  \bibitem{} Ellingsen, S. P., von Bibra, M. L., McCulloch, P. M., Norris, R. P., Deshpande, A. A., \& Phillips, C. J. 1996, \mnras, 280, 378
  \bibitem{} Harju, J., Lehtinen, K., Booth, R. S., \& Zinchenko, I., 1998, \aaps, 132, 211
  \bibitem{} Hasegawa, T. I., \& Mitchell, G. F. 1995, \apj, 451, 225
  \bibitem{} Kurtz, S., Churchwell, E., \& Wood, D. O. S., 1994, \apjs, 91, 659
  \bibitem{} Malyshev, A. V., \& Sobolev, A. M., 2003, A\&AT, 22, 1
  \bibitem{} Marengo, M., Jayawardhana, R., \& Fazio, G. G. et al., 2000, \apj, 541, L63
  \bibitem{} Minier, V., Ellingsen, S. P., Norris, R. P., \& Booth, R. S., 2003, \aap, 403, 1095
  \bibitem{} Minchin, N. R., White, G. J., \& Padman, R, 1993, \aap, 277, 595
  \bibitem{} Molinari, S., Testi, L., Rodr$\acute{i}$guez, L. F., \& Zhang, Q., 2002, \apj, 570, 758
  \bibitem{} Ott, M., Witzel, A., \& Quirrenbach, A., 1994, \aap, 284, 331
  \bibitem{} Palagi, F., Cesaroni, R., and Comoretto, G. et al, 1993, \aaps, 101, 153
  \bibitem{} Palla F., Brand J., Cesaroni R. et al., 1991, A\&A 246, 249
  \bibitem{} Pandian, J. D., \& Goldsmith, P. F. 2007, \apj, 669, 435
  \bibitem{} Pandian, J. D., Goldsmith, P. F., \& Deshpande, A. A. 2007, \apj, 656, 255
  \bibitem{} Pestalozzi M., Minier V. \& Booth R. 2005, \aap, 432, 737
  \bibitem{} Sobolev, A. M., Ostrovskii, A. B., Kirsanova, M. S. et al. 2005, IAUS, 227, 174
  \bibitem{} Szymczak, M., Pillai, T., \& Menten, K. M., 2005, \aap, 434, 613
  \bibitem{} Szymczak, M., Kus, A. J., Hrynek, G., Kepa, A., \& Pazderski, E. 2002, \aap, 392, 277
  \bibitem{} Szymczak, M., Hrynek, G., \& Kus, A. J., 2000, \aaps, 143, 269
  \bibitem{} Trinidad, M. A., Curiel, S., Cant$\acute{o}$, J. et al., 2003, \apj, 589, 386
  \bibitem{} van der Tak F., van Dishoeck E. F., Evans II N.J. et al, 1999, \apj, 522, 991
  \bibitem{} van der Walt, J., 2005, \mnras, 360, 153
  \bibitem{} Walsh, A. J., Hyland, A. R., Robinson, G., \& Burton, M. G, 1997, \mnras, 291, 261
  \bibitem{} Wood, D. O. S., \& Churchwell, E, 1989a, \apjs, 69, 831
  \bibitem{} Wood, D. O. S., \& Churchwell, E, 1989b, \apj, 340, 265
  \bibitem{} Wouterloot, J. G. A., Brand, J., 1989, \aaps, 80, 149
  \bibitem{} Wouterloot, J. G. A., Brand, J., Fiegle, K., 1993, \aaps, 98, 589
  \bibitem{} Wouterloot, J. G. A., \& Walmsley, C. M., 1986, A\&A, 168, 237
  \bibitem{} Wu, Y., Zhang, Q., Yu, W. et al., 2006, \aap, 450, 607
  \bibitem{} Xu, Y., Reid, M. J., Zheng, X. W., \& Menten, K. M. 2006, Science, 311, 54
  \bibitem{} Xu, Y., Zheng X. W., \& Jiang, D. R. 2003, \cjaa, 3, 49

\end{thebibliography}
\end{document}